\documentclass[aps,superscriptaddress,twocolumn,amssymb,showpacs,prl]{revtex4-1}  
\usepackage{graphicx}
\usepackage{epsfig}
\pdfoutput=1
\usepackage{bm}
\usepackage{amsmath}
\usepackage{color}

\begin{document}
\title{A New Approach to Phonon Anharmonicity}

\author{Krzysztof Parlinski  }
\affiliation{Institute of Nuclear Physics, Polish Academy of Sciences,
             Radzikowskiego 152, PL-31342 Krak\'ow, Poland}
\affiliation{Computing for Materials, Krak\'ow, Poland}

\begin{abstract}
An approach  to compute the anharmonic peaks of the phonon dispersion 
curves through the {\it ab initio} calculated Hellmann-Feynman forces 
from a series of supercells with realistic atomic displacements
of all atoms,
which correspond to a given temperature, is reported. Obtained phonon 
dispersion bands are able to represent the positions and shapes 
of the anharmonic peaks.
As example, the approach to cubic PbTe and perovskite 
MgSiO$_3$ crystals is applied.
\end{abstract}

\date{\today}

\maketitle

\par The conventional treatment of anharmonicity in crystals relies on the 
expansion of the potential energy with respect to atomic displacements
up to orders higher then quadratic terms.
Typically, third and fourth orders
expansion terms are used \cite{cowley}, 
in spite of existence of powerful perturbation 
methods being able to sum part of expansion diagrams \cite{amaradudin,barron}. 
Unfortunately, these methods are computational very expensive.

\par
The anharmonicity is required for derivation  of
the thermal conductivity of materials,
specially thermoelectrics, which efficiency to transfer heat to 
electricity increases with lowering  the heat transfer.
Moreover, the high pressures  thermal conductivity is important to understand 
heat transfer in the Earth interior.

\par 
In this Letter a new approach to handle crystals anharmonicity is proposed.
It relies in probing the potential energy landscape of the crystal
with atomic vibration.
Then, the vibrations are transformed and classified in 
the wavevector reciprocal space, similar to harmonic case.
The expansion of the potential energy is not needed. 

\par
The existing density functional theory (DFT) codes provide sufficiently
accurate Hellmann-Feynman (HF) forces to calculate the phonon 
frequencies and properly describe related phenomena.
The DFT approach requires to approximate the crystal as a  {\it supercell},
which is a parallelepiped construction made of primitive unit cells.
By definition periodic boundary conditions are imposed on the supercell.

\par
To compute harmonic phonons, the {\it ab initio} force constant approach
formulated by Parlinski, Li,  Kawazoe, in 1997 \cite{prlparlinski}, 
can be used.
This method has been already  applied  for several hundreds of crystals
and the results show good agreement with the measured phonon data.
The computing method relies in using the relation between harmonic forces 
${\bf F_H}({\bf n}, \mu)$,
induced by atoms displaced by
${\bf u}({\bf m}, \nu)$ from the equilibrium position
\begin{equation}
{\bf F_H}({\bf n},\mu ) = -
\sum _{{\bf m},\nu }
{\bf B_H}({\bf n},\mu ,{\bf m},\nu ) \cdot {\bf u}({\bf m},\nu )
\label{forces}
\end{equation}
where ${\bf n}$, ${\bf m}$ label the primitive unit cells 
in the supercell, 
and $\mu$, $\nu$ number atoms within these cells. 
The ${\bf B_H}({\bf n},\mu ,{\bf m},\nu )$
are harmonic force constants between marked atoms and
they determine the conventional dynamical matrix 
$D({\bf k}, \mu,\nu)=\frac{1}{\sqrt{M_{\mu}M_{\nu}}}
\sum _{{\bf m}}
{\bf B_H}(0,\mu ,{\bf m},\nu ) 
exp\{-2\pi{\bf k}\cdot \{{\bf R}(0,\mu )-{\bf R}({\bf m},\nu)]\}$
\cite{maradudin}.
The eigenvalue equation 
$\omega ^2({\bf k},j){\cal E}({\bf k},j) =
{\bf D}({\bf k}){\cal E}({\bf k},j)$
of this Hermitian matrix provides phonon frequencies
$\omega ^2({\bf k},j)$,  and eigenvectors
${\cal E}^{(0))}({\bf k},j)$.
In Ref. \cite {prlparlinski} was proposed to decouple 
each force constant matrix into a product
\begin{equation}
{\bf B_H}({\bf n},\mu ;{\bf m},\nu ) =
         {\bf A}({\bf n},\mu ;{\bf m},\nu )
         \cdot
         {\bf P_H}({\bf n},\mu ;{\bf m},\nu )
\label{decouple}
\end{equation}
where $(3\times 3)$ ${\bf B_H}({\bf n},\mu ,{\bf m},\nu )$ matrix
is rearranged to $(9\times 1)$ matrix,
${\bf A}({\bf n},\mu ;{\bf m},\nu )$ is a $(9\times p)$
matrix, and
${\bf P_H}({\bf n},\mu ;{\bf m},\nu )$ is a $(p\times 1)$ matrix.
The ${\bf A}({\bf n},\mu ;{\bf m},\nu )$ is entirely 
determined by crystal symmetry, and does 
not involve potential parameters.
The parameter matrix 
${\bf P_H}({\bf n},\mu ;{\bf m},\nu )$
depends on potential details only, and index $p$
denotes number of independent parameters of the force constant in question.
Rearranging the three components of displacement vector  
${\bf u}({\bf m}, \nu)$ into $(3\times9)$ matrix 
${\bf U}({\bf m}, \nu)$, 
Eq.(\ref{forces}) becomes
\begin{equation}
{\bf F_H}({\bf n},\mu ) = 
\sum _{{\bf m},\nu ,j}
{\bf C_U}({\bf n},\mu ,{\bf m},\nu )
\cdot
{\bf P_H}({\bf n},\mu ;{\bf m},\nu )
\label{harmaster}
\end{equation}
where the $(9\times p)$ {\it symmetry adapted displacement matrix} 
${\bf C_U}({\bf n},\mu ,{\bf m},\nu )$ 
has been determined as
\begin{equation}
{\bf C_U}({\bf n},\mu ;{\bf m},\nu ) =
         - {\bf U}({\bf m}, \nu)
         \cdot
         {\bf A}({\bf n},\mu ;{\bf m},\nu ).
\label{notation}
\end{equation}
The ${\bf C_U}({\bf n},\mu ;{\bf m},\nu )$
matrix is known from used atomic displacements and crystal symmetry.
For small atomic displacements of order of 0.03\AA ,
the harmonic forces ${\bf F_H}({\bf n},\mu )$  can be obtained from 
{\it ab initio} calculated HF forces. 
These data permit to find all unknown 
independent parameters ${\bf P_H}({\bf n},\mu ;{\bf m},\nu )$, and from 
Eq. (\ref{decouple}), all force constants 
${\bf B_H}({\bf n},\mu ;{\bf m},\nu )$
within the supercell.

\par
Single atomic displacements in the primitive unit cell $(0, \nu )$
is sufficient to create single HF force field for that atom.
A force field is a set of $3n$ HF forces obtained for the run 
in {\it ab initio} code with 
a single displaced atom. Here, $n$ denotes the number of atoms in the supercell.
Usually a single HF force field is not sufficient. A minimum  number $s$ of 
HF force fields 
is equal to the number of non-equivalent atoms of primitive 
unit cell, supplemented by a number of non-equivalent directions 
of the displacements. Finally we have $3ns$ equation of kind (\ref{harmaster})
to be solved simultaneously. Collecting these $3ns$ equations to a
{\it global form}, the system of equations for the harmonic force constant
parameters can be written as
\begin{equation}
{\cal F_H} =  {\cal C_U}  \cdot {\cal P_H}
\label{HMaster}
\end{equation}
where ${\bf{\cal F_H}}$, ${\bf {\cal C_U}}$ and  ${\bf{\cal P_H}}$ are
$(3ns\times 1)$, $(3ns\times p')$ and $(p'\times 1)$ dimensional matrices,
respectively,
where $p'$ is now the total number of independent parameters needed for 
all force constants within a supercell.
In this system of equations the number of HF forces is greater 
then the number of potential parameters, $3ns > p'$,
therefore, this is an {\it overdetermined} system.
To solve it, the singular value decomposition method (SVD) 
\cite{numericalrecipies} to matrix 
${\bf {\cal C_U}}$ is applied \cite{prlparlinski}. 
Then the potential parameters
can be found from HF force 
${\cal P_H} = {\cal C_U}^{-1}\cdot {\cal F_H}$.
This SVD method provides a solution, which is the best approximation 
in the least square sense.
The above procedure delivers very effective method to find harmonic 
phonons, their frequencies, polarization vectors, phonon dispersion curves, 
phonon density of states, and many other phonon dependent quantities.
 
\par 
The global form of system of equations,  Eq. (\ref{HMaster}),
permits to add other conditions to be fulfilled.
The simple example are the translational-rotational invariants
\cite{maradudin}, which can be reformulated to a form of matrix 
${\bf{\cal M}}$ of $(18n \times p')$ dimensions
using Eq.(\ref{decouple}). Then, Eq.(\ref{HMaster}) becomes
\begin{equation}
\left(
  \begin{array}{c}
    {\cal F_H}\\
          0 \\
  \end{array}
\right)  =
\left(
  \begin{array}{c}
    {\cal C_U}\\
    \beta {\cal M} \\
  \end{array}
\right)
\cdot
{\cal P_H}
\label{tranrot}
\end{equation}
where $\beta$ is adjusting the strength to satisfy the translational-rotational conditions.
Eq. (\ref{tranrot}) is solved by SVD method.

\par
To compute anharmonic effects one may use a similar approach.
Any DTF calculations of HF forces contain 
an information on the anharmonicity.
This means that there is an access to anharmonic landscape of the 
crystal potential energy. Generally, a HF force 
${\bf F}({\bf n},\mu )$ acting on an atom $({\bf n}, \mu)$ 
can be  treated as originating either from harmonic  
${\bf F_H}({\bf n},\mu )$, or from
anharmonic ${\bf F_A}({\bf n},\mu )$ contributions. 
In the pure harmonic regime the matrices 
${\bf B_H}({\bf n},\mu ;{\bf m},\nu )$,
Eq.(\ref{forces}), do not depend on the amplitude of atomic displacements, 
hence they are temperature independent.
It follows from linear dependence between harmonic forces and
atom displacements.

\par
The anharmonic potential has a many-body character, involving 
at a given temperature a lot of 
atoms simultaneously displaced.
Additionally,  all atoms of the crystal move in time, 
varying the displacements around the equilibrium position, and
at a given time moment all atoms of the supercell 
create pattern of displacement ${\bf U^{(i)}}$, denoted as  $(i)$.
One may treat also the displacement pattern $(i)$ as coming at 
the same time from different
locations of the crystal.
Each displacement pattern $(i)$ allows to calculate one HF 
anharmonic force field 
${\bf F_A^{(i)}}$. Knowing the displacement patterns ${\bf U^{(i)}}$
and force fields ${\bf F_A^{(i)}}$, we supplement the Eq.(\ref{tranrot})
by anharmonic contributions,
formulating in this way {\it a  global anharmonic form} of system of equations
\begin{equation}
\left(
  \begin{array}{c}
    {\cal F_H}\\
          0 \\
    {\cal F_A}^{(i)} \\
  \end{array}
\right)  =
\left(
  \begin{array}{c}
    {\cal C_U}\\
    \beta {\cal M} \\
    {\cal C}_{U^{(i)}}\\
  \end{array}
\right)
\cdot
{\cal P_A}^{(i)}
\label{anharmasteri}
\end{equation}
where $i=1, 2, \dots  N$ runs over displacement patterns.

\par
The SVD solution of Eq.(\ref{anharmasteri}), corresponding to single 
displacement pattern, gives a single set of 
dispersion curves of index $(i)$.
Since in space, or in time the displacement pattern ${\bf U^{(i)}}$
changes, for a realistic modelling of anharmonic effects
one should select a sequence of N such patterns 
($i=1,2,\dots  N$), and compute dispersion curves for 
all these $N$ patterns. From set to set, probing different 
anharmonic environments, 
the phonon dispersion curves are slightly different due to 
variation of potential energy reached now by displaced atoms.
Indeed, in the absence of anharmonicity, 
for example with negligible amplitudes of atomic displacements,
all sets $(i)$ of phonon dispersion 
curves will be identical to harmonic case within the accuracy of 
a computational noise and approximate SVD solution.
In anharmonic case all phonon dispersion curves, $i=1,2,\dots   N$
form  bands, which reflect the anharmonic character of the potential
energy landscape, caused by the atomic displacements around the 
global potential minima.

\par
To stay consistent with the supercell concept, the displacement 
patterns ${\bf U^{(i)}}$ must preserve the periodic boundary conditions.
Thus, a displacement pattern must be a superposition of the allowed
phonon displacement waves.
Hence, each phonon displacement wave 
must be commensurate with the supercell size and shape. 
Its amplitude can be estimated from the harmonic theory as
\begin{equation}
{\bf U}({\bf n},\mu) = \frac{Q({\bf k},j)}{\sqrt{M(\mu)}} 
          exp[2\pi i({\bf k}\cdot {\bf R}({\bf n}, \mu) - \phi ({\bf k}, j)]
\label{wave}
\end{equation}
where the mean square amplitude of the wave is determined by 
\begin{equation}
<Q^2({\bf k},j)>  =  
\frac{\hbar  }{2\omega ({\bf k},j)} coth\left(\frac{\hbar \omega({\bf k},j)}
{2k_BT} \right)
\label{amplitude}
\end{equation}
and $\omega ({\bf k},j)$ are the harmonic phonon frequencies
at wavevector ${\bf k}$, and phonon branch $j$. The phase factor 
$\phi({\bf k},j)$ could be taken at random to mimic different
displacement patterns. The temperature displacement variation,
Eq.(\ref{wave}) depends on the thermal occupation factor, Eq.(\ref{amplitude}).
The HF forces calculated for a displacement pattern include 
contributions arising from many atoms displaced simultaneously.
The approximation relies on using only displacement waves of 
a few hundreds discreet wavevectors.
We remark that the displacement patterns may also be obtained
in another way, for example, as snapshots of atomic motion traced 
during molecular dynamic simulation. But even in this case 
the confinements imposed by selecting only commensurate displacement 
waves holds as well.


\par
The crystal symmetry is determined by the crystallographic space group.
In harmonic regime all fluctuations - atomic displacements - are
described and classified by the phonon modes.
To each phonon mode an irreducible representation of the crystal
space group is assigned.
The above classification applies to phonon modes derived from 
force constants being a solution of Eq.(\ref{tranrot}), and each set $(i)$
of global system of equations, Eq.(\ref{anharmasteri}).
Then, it follows that the current 
anharmonic phonon bands are characterized by the same
irreducible representations as the irreducible representations of 
corresponding phonon dispersion relations of the harmonic case.
However, the anharmonic phonon bands have much larger chances to overlap.
Therefore, a systematic method to select out a single anharmonic
peak from phonon bands is required.

\par Now the {\it projection method} to select out a particular anharmonic mode
is proposed. The conventional assignment of irreducible representations 
to harmonic phonons is considered as a reference to classify
the anharmonic peaks. Then, diagonalization of the dynamical matrix, 
delivers orthonormalized eigenvectors 
${\cal E}^{(0)}({\bf k},J)$, and ${\cal E}^{(i)}({\bf k},j)$
of harmonic phonons curves and anharmonic phonon bands 
relevant for all displacement 
patterns $(i)$, where $J$ and $j$ label phonon modes for the same
wavevector ${\bf k}$, respectively.
Each eigenvector involved in the anharmonic peak can be expanded over
complete set $J=1, 2,\dots J_{max}$ of harmonic eigenvectors
\begin{equation}
{\cal E}^{(i)}({\bf k},j ) =
\sum _{J=1}^{J_{max}}
\alpha ^{(i)}({\bf k},j,J) {\cal E}^{(0)}({\bf k},J ).
\label{expansion}
\end{equation}
Applying the orthonormality relation
$\sum _{{\bf j}=1}^{j_{max}}  
{\cal E}^{*(i)}({\bf k},j )\cdot {\cal E}^{(i)}({\bf k},j) = 1$,
the expansion coefficients of Eq.(\ref{expansion}) can be found as
$\alpha ^{(i)}({\bf k},j,J)= {\cal E}^{*(i)}({\bf k},j)
                       \cdot {\cal E}^{(0)}({\bf k},J)$.
At fixed ${\bf k}$ wavevector {\it the anharmonic $J$ mode of phonon
density of states}, 
denoted by $b_J(\omega , {\bf k})$, as a function of frequency $\omega $,
can be found from the histogram
\begin{equation}
b_J(\omega , {\bf k}) =
\frac{1}{Z}
\sum _{i=1}^{N} \sum _{j=1}^{j_{max}} 
\mid \alpha ^{(i)}({\bf k},j,J)\mid ^2 
\delta _{\Delta \omega}(\omega - \omega ^{(i))}({\bf k},j))
\label{density}
\end{equation}
where $ Z=N\cdot J_{max}\cdot j_{max}\cdot \Delta \omega$.
The histogram bin $\Delta \omega$ is defined by the function 
$\delta _{\Delta \omega}(x)=1$, if 
$-\frac{\Delta \omega}{2} < x \le \frac{\Delta \omega}{2}$,
or $0$ otherwise.
The summation $i=1, 2, \dots N$ runs over
all displacement patterns $(i)$.
The coefficients $\alpha ^{(i)}({\bf k},j,J)$ select out from
all $(i)$ bands only those phonon intensities which resemble
the vibrations determined by harmonic eigenvectors
${\cal E}^{(0)}({\bf k},J)$.
Each anharmonic mode of phonon density of states $b_J(\omega , 
{\bf k})$ determines 
a single anharmonic mode of symmetry $J$.
The referred method is able to select out all single 
anharmonic peaks, even if they overlap.

\begin{figure}[t!]
\includegraphics[width=8.4cm]{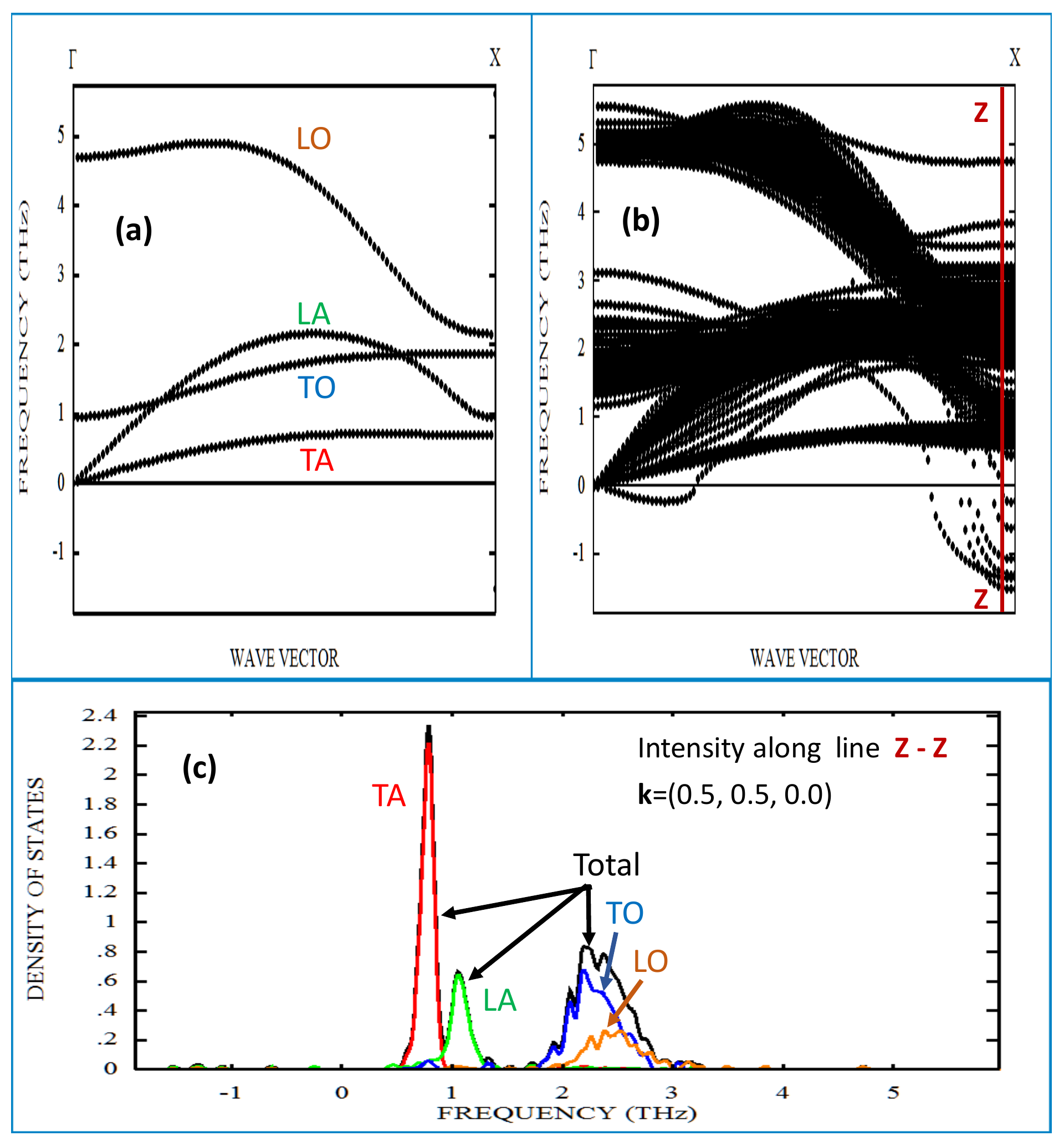}
\caption{(Color online) 
The cubic PbTe phonon dispersion relations along $\Gamma (0,0,0) - X(\frac{1}{2},0,0)$
wavevector path.
(a) Harmonic curves at $T=0K$ (temperatureless regime), and pressure $P=0GPa$.
(b) Anharmonic phonon bands originated from 100 different sets of 
displacement patterns at $T=300K$ and $P=0.828GPa$.
(c) Anharmonic mode density of states of all phonons, Eq(\ref{density}),
for a fixed wavevector ${\bf k} = X$ (vertical z-z line at (b) plot), and
as a function of phonon frequency $\omega$ in $THz$.
The total spectrum was separated, Eq.(\ref{density}), to TA, LA, TO and LA anharmonic peaks.}
\label{fig:Fig1}
\end{figure}

\par
The above approach is illustrated with two examples:
PbTe and MgSiO$_3$ crystals.
Our intention, however, is not to provide a deep physical analysis
of the samples, 
but to show what kind of results can be obtained.
All numerical calculations were performed within DFT method using 
VASP code \cite{vasp}, and applying GGA-PAW approach issued with
this code. Phonons were calculated with  PHONON code \cite{prlparlinski}.
Displacements of $0.03$ $\AA$ were used to generate the lists of 
harmonic HF forces. 

\par
The anharmonicity of cubic PbTe semiconductor  
has been extensively studied in Ref. \cite{delaire}.
For PbTe, the space group $Fm{\bar 3}m$, $2\times 2\times 2$ supercell
with $64$ atoms, and
$4\times 4\times 4$ $k$-point mesh were used.
The optimized lattice constant for pressure $P=0GPa$ was $a = 6.557 \AA$. 
The effective charges of $Z^* = \pm 5.8$ and electronic dielectric constant
$\epsilon = 25.26$ were taken from Ref.\cite{zhang}.
The harmonic phonon dispersion relations are shown on Fig.\ref{fig:Fig1}a
along $\Gamma (0,0,0) - X(\frac{1}{2},0,0)$ wavevector path.
The computed eigenvectors specify uniquely transverse (T),
longitudinal (L), acoustic (A), and optical (O) polarizations,
indicated as TA, LA, TO, LO modes on the plot. 

\par
For temperature $T=300K$,
100 displacement patterns were generated using in each case $189$ phonon waves 
with wavelengths commensurate with $2\times 2\times2$ supercell.
For each pattern the phonon phase was taken at random.
The average atomic displacements over all patterns were 0.16 and 0.12 $\AA$,
and from analytical Eq.(\ref{amplitude},) with harmonic density of states
were 0.15 and 0.12 $\AA$, 
for Pb and Te, respectively. 
Displaced atoms increase the potential 
energy of the pattern with respect to $T=0K$ state 
by $\Delta E$. Average value of $\Delta E$, 
calculated from the harmonic form of the potential energy derived using
the same force constant as involved in phonon bands, and on other hand
from the energy excess
provided by VASP, were pretty close, and read 75.37 and 74.13
meV/prim.unit cell, respectively. Moreover, the displaced patterns,
according to VASP, 
created a pressure of 0.828 GPa in the supercell with the same lattice constant 
as in $T=0K$, namely $a = 6.557 \AA$ . It is natural since the calculation 
at $T=0K$ and $T=300k$ were 
done at the same supercell volume.

\par
Fig.\ref{fig:Fig1}b shows plots of phonon bands calculated 
for 100 displacement patterns. 
Each band consists of 100 phonon curves (200 curves for doubly degenerate modes)
of the same symmetry. There are regions were bands overlap.
Having 600 modes for each wavevector ${\bf k}$ one may construct a histogram
of the anharmonic mode phonon density of states. 
Such a histogram, plotted for a single
wavevector ${\bf k} = X(\frac{1}{2},0,0)$ (line z-z), 
is presented on Fig.\ref{fig:Fig1}c.
It shows total mode phonon density of states and its separation into four 
specific anharmonic modes TA, LA, TO, and LO.
The TA and LA modes 
look rather sharp, while TO and LO are much wider.  
Since the shape of each 
anharmonic peak is known, there is no problem to establish peak 
position and width.
Without any additional {\it ab initio} calculations,
but for the same $T$ and $P$,
the anharmonic peaks for any wavevector ${\bf k}$ can be found.
One notices that on Fig.\ref{fig:Fig1}b imaginary (negative) 
phonon branches appear, however,
they are very rare, and their contributions to anharmonic mode 
of phonon density of states,  Fig.\ref{fig:Fig1}c, is negligible.

\begin{figure}[t!]
\includegraphics[width=8.4cm]{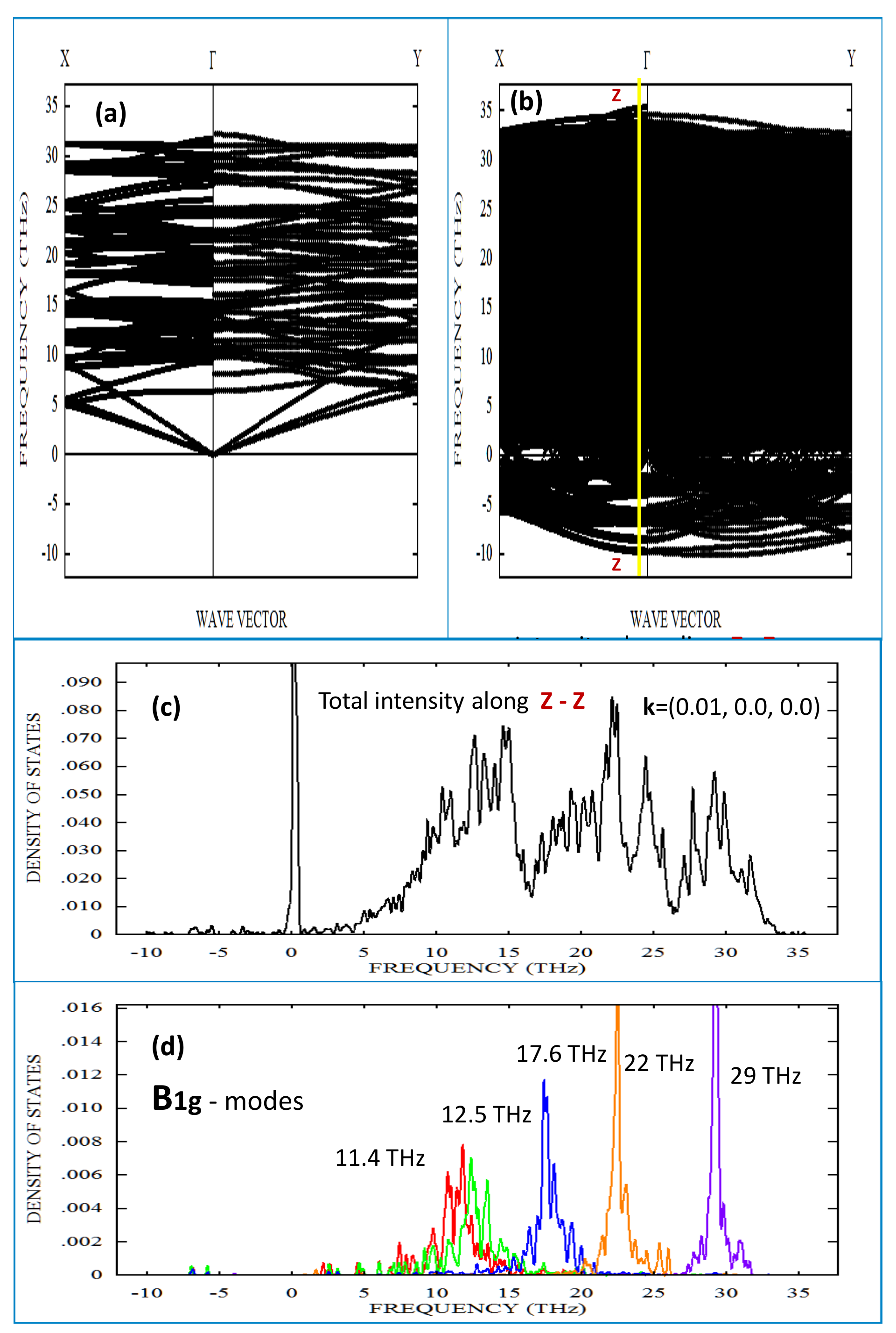}
\caption{(Color online)
The orthorhombic perovskite MgSiO$\,_3$ 
phonon dispersion curves along 
$X(\frac{1}{2},0,0) - \Gamma (0,0,0) - Y(0,\frac{1}{2},0)$
wavevector path.
(a) Harmonic dispersion relations at $T=0K$ 
(temperatureless regime), and pressure $P=57.3GPa$.
(b) Anharmonic phonon bands originated from 100 different sets of 
displacement patterns at $T=2300K$ and $P=70.5GPa$.
(c) Anharmonic mode density of states of all phonons, Eq(\ref{density}),
for a fixed wavevector close to  ${\bf k} = \Gamma (0,0,0))$
(vertical z-z line at (b) plot), 
as a function of phonon frequency $\omega$ in $THz$.
(d) Plot of all anharmonic peaks of $B_{1g}$ symmetry
at $\Gamma$ selected out using Eq.(\ref{density}),
from complete anharmonic mode density of states (line z-z at (b))plot)
at ${\bf k} = \Gamma(0,0,0)$.}
\label{fig:Fig2}
\end{figure}

\par
The anharmonicity of MgSiO$_3$ has already been
carried on with respect to thermodynamical properties and
thermal conductivity \cite{dekura} of the  Earth lower mantle,
where high $T$ and $P$ persist.
The perovskite MgSiO$_3$ belongs to orthorhombic space group $Pmnb$, No=62.
For all runs the 
$1\times \sqrt{2} \times \sqrt{2}$ supercell, and $2\times 2\times2$
$k$-point mesh were selected. For effective charges the formal charges
were chosen.
The anharmonic calculations started from  optimization of the supercell at pressure of
$P=57.3GPa$ and  $T=0K$. Hence, the lattice constants
$a = 6.481\AA $,  $b = 4.689\AA $,  $c = 4.462\AA $ were found. 
The harmonic phonon dispersion relations were plotted along 
$X(\frac{1}{2},0,0) - \Gamma (0,0,0) - Y(0,\frac{1}{2},0)$
wavevector path, and are shown on Fig.\ref{fig:Fig2}a.  
Since the primitive unit cell contains 20 atoms, 
therefore there appears 60 phonon dispersion curves.

\par
For the anharmonic analysis temperature of $T=2300K$ was chosen,
and  100 displacement patterns was generated. 
Each pattern used 117 phonon waves with wavelength 
commensurate to the supercell size.
The average atomic displacements 
over all patterns were slightly 
anisotropic, but the average values of them were: 
0.125, 0.105, 0.128, and 0.089 $\AA$,
and from analytical Eq.(\ref{amplitude},) with harmonic density of states
were 0.110, 0.105, 0.127 and 0.087 $\AA$, 
for O1, O2, Mg and Si atoms , respectively.
The potential energy increased 
due to presence of the atomic displacements. 
The energy excess $\Delta E$ calculated directly from 
force constants and from VASP were  
6.049 and 6.727 eV/primitive unit cell, respectively.
The displacement patters created the pressure $P=70.5GPa$
for the same volume as used in $T=0K$.

\par
Next, the anharmonic properties  for $T=2300K$
and $P=70.5GPa$ were calculated.
Fig.\ref{fig:Fig2}b shows plots of phonon bands calculated along 
similar wavevector path as for harmonic phonon curves.
Each band consists of 100 phonon curves
of the same symmetry. There is so many bands that they overlap,
and the information becomes obscure.
Moreover, there are observed phonon curves which exists 
in the imaginary frequency region,
although  their intensities were negligible.
Having 6000 modes for each wavevector ${\bf k}$ one may construct a histogram
of the total anharmonic mode phonon density of states. 
Such a histogram, plotted for a single
wavevector close to ${\bf k} = \Gamma(0,0,0)$ (line z-z at (b) plot), 
is presented on Fig.\ref{fig:Fig1}c. 
Applying the projection method, Eqs (\ref{expansion},\ref{density}) 
to the total anharmonic mode phonon density of states,
the separation of all 60 anharmonic peaks 
could be done. 
For example, choosing from 60 $\Gamma$-modes
$7A_g + 5B_{1g} + 5B_{2g} + 7B_{3g} +  8A_u + 10B_{1u} + 10B_{2u} + 8B_{3u}$
one kind, for example $B_{1g}$ and projecting the anharmonic eigenvectors
onto harmonic eigenvectors, the $B_{1g}$ anharmonic peaks were found and
plotted on Fig.\ref{fig:Fig2}d. All peaks are well defined, and could be used to
determine their positions and widths, and hence the phonon lifetimes. 
Such a procedure of selection anharmonic modes can be applied to any wavevector.

\par
In conclusions we turn attention on several aspects:
(i) the phonon lifetime $\tau ({\bf k},J)$
is usually found from the width of the anharmonic peak 
$\tau ^{-1}({\bf k},J) = width _{\omega} (b_J(\omega ,{\bf k})$,
where operator $width_{\omega }$ denotes a procedure to elucidate 
the width from fitted Gaussian, Lorentzian, or similar function.
Some of the peaks may contain low intensity tails, which influence 
the phonon lifetime.
(ii) the {\it ab initio} code typically is handling phonon-phonon, 
electron-phonon, and magnetic-phonon interactions, and these effects are 
included in the present method. 
(iii) the anharmonicity is measured by 
spectroscopic methods. The description of the cross sections are as a rule
supplemented by certain form factor, and corresponding theoretical
expression can be derived for phonon bands. It means that the 
expected spectroscopic spectra can be determined from phonon bands.
(iv) for the Raman scattering, or infrared absorption
the current method offer to include changes of 
Raman tensor itself, or effective charges itself due to atomic displacements in 
the patterns,
respectively. 
This effects might influence the  Raman or infrared spectra.
(v) anharmonic displacement patterns offer full parallelization,
The CPU time on the cluster for MgSiO$_3$ example was about 10 hours only.

The author acknowledges discussions with P.Piekarz, A.Ole\'s, U.Wdowik, J.Jochym,
M.Sternik and J.\L{}a\.zewski who also helped to run some calculations.

\end{document}